\documentclass[a4paper, oneside, twocolumn, notitlepage, 10pt]{extarticle_ecoc}
\usepackage{ecoc}

\usepackage[table]{xcolor}

\begin{document}
\selectlanguage{english}    


\title{Tx-Rx Mode Mismatch Effects in Gaussian-Modulated CV QKD}%


\author{
    Mateusz Kucharczyk\textsuperscript{(1,2)}, Micha{l} Jachura\textsuperscript{(1)},
    Marcin Jarzyna\textsuperscript{(1,3)}, Konrad Banaszek\textsuperscript{(1,2)},\\
    Amirhossein Ghazisaeidi\textsuperscript{(4)}
}

\maketitle                  


\begin{strip}
    \begin{author_descr}

        \textsuperscript{(1)} Centre for Quantum Optical Technologies, CeNT, University of Warsaw, 
   02-097 Warszawa, Poland,
        \textcolor{blue}{\uline{m.jachura@cent.uw.edu.pl}}

        \textsuperscript{(2)} Faculty of Physics, University of Warsaw, Pasteura 5, 02-093 Warsaw, Poland

        \textsuperscript{(3)} Department of Optics, Palack\'{y} University, 17. listopadu 1192/12, 771 46 Olomouc, Czech Republic

    \textsuperscript{(4)} Nokia Bell Labs, 91300 Massy, France

    \end{author_descr}
\end{strip}

\renewcommand\footnotemark{}
\renewcommand\footnoterule{}


\begin{strip}
    \begin{ecoc_abstract}
The impact of technical limitations on pulse shaping used to generate a CV QKD signal is quantified in terms of the attainable secure key rate. Optimization of key spectral efficiency for Gaussian-modulated CV QKD  with truncated and discretized root-raised cosine profiles is discussed. 
\textcopyright2024 The Author(s)
    \end{ecoc_abstract}
\end{strip}


\section{Introduction}
Quantum key distribution (QKD) is being explored as a physical layer security solution in optical communication systems that could be made robust even against the most sophisticated adversarial attacks involving a prospective use of quantum computers\cite{LaudenbachAQT2018,PirandolaAOP2020}. Among many QKD options, continuous-variable (CV) protocols avoid the need for single-photon detection technology and may be implemented with components and subsystems developed for high-speed coherent optical communication, which eventually should enable operation of CV QKD systems at multi-gigabaud symbol rates\cite{RoumestanOFC2021,RoumestanECOC2021,PanWangOL2022}. However, the intrinsically different nature of the key generation task compared to regular information transmission requires careful analysis of the performance of coherent communication techniques in QKD applications with respect to the attainable key rates and security \cite{Jouguet2013, Hao2016, Hajomer2022}.

The purpose of this contribution is to discuss theoretically, as a case study, the effects of limited pulse shaping capability in Gaussian-modulated CV QKD. Specifically, we consider a CV QKD signal in the form of root-raised cosine (RRC) pulses generated using a finite number of discrete samples (taps). It is shown that the mode mismatch between such pulses and actual RRC profiles detected by the receiver results in intersymbol interference (ISI) \cite{Armstrong2009}, whose impact on the attainable secret key rate and the key spectral efficiency is analyzed in quantitative terms. 

\section{CV QKD Security}
\begin{figure}[b]
    \centering
    \includegraphics[width=7.5cm]{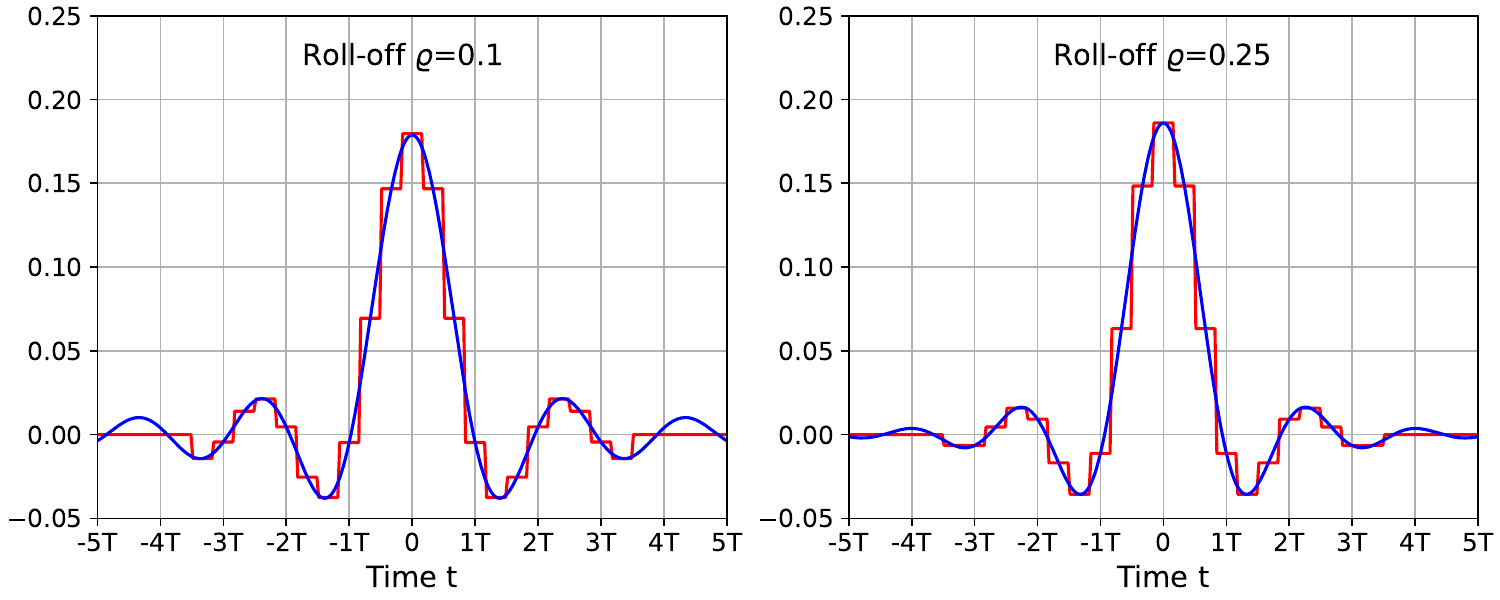}
    \caption{Root-raised cosine profiles (blue) and their truncated sample-and-hold approximations
    for the CV QKD signal (red).}
    \label{Fig:RRCprofiles}
\end{figure}

The standard scenario examined in most CV QKD security analyses is that Alice's transmitter Tx sends a discrete-time sequence of complex electromagnetic field amplitudes (symbols) $\alpha_j$, $j=\ldots,-1,0,1,\ldots$ whose values are subsequently detected by Bob by means of coherent detection. For concreteness, Bob's receiver Rx will be taken  in the form of an optical hybrid measuring jointly both $I$ and $Q$ field quadratures. The scenario presented above implicitly assumes that the temporal mode (pulse shape) for each symbol is orthogonal in terms of the scalar product to all others and therefore can be detected separately by Bob. For the $j$th symbol prepared in a normalized complex-baseband mode $v(t-jT)$, where $T$ is the symbol temporal spacing, the orthogonality assumption is expressed mathematically as\cite{BanaszekJLT2020}
\begin{equation}
\int_{-\infty}^{\infty} \mathrm{d}t \, v^\ast(t-jT) v (t-j'T) = \delta_{jj'} 
\label{Eq:Orthogonality}
\end{equation}
for any integers $j, j'$.
The above requirement can be equivalently written in terms of the spectral amplitude  $V(f) = \int_{-\infty}^{\infty} \textrm{d}t \, {\textrm e}^{-2\pi \mathrm{i} ft} v(t)$ in the form:
\begin{equation}
\int_{-\infty}^{\infty} \mathrm{d}f \, \mathrm{e}^{2\pi \mathrm{i} (j-j') f T}
|V(f)|^2 = \delta_{jj'}
\end{equation}
which is instantly recognized as the no-ISI condition for the spectral pulse power profile $|V(f)|^2$. Importantly, this condition is satisfied by the class of RRC profiles with any roll-off factor $0 \le \varrho \le 1$, depicted in Fig.~\ref{Fig:RRCprofiles} for two exemplary values of $\varrho$. 

In Gaussian-modulated CV QKD, individual symbol values are i.i.d.\ random variables drawn from a Gaussian ensemble $\alpha_j \sim {\mathcal{CN}} (0, \bar{n})$. It is convenient to choose units such that $\bar{n}$ specifies the average photon number per symbol. Transmission of the QKD signal through an optical channel is characterized by the power transmission coefficient $\tau$ and the excess noise power spectral density $n_\mathrm{n}$ specified in photon number units at the channel output. The standard, most conservative scenario considered in QKD security analyses is that any deviations from the ideal lossless and noiseless channel result from actions of an adversary, Eve. For an additive Gaussian excess noise the attainable secret key rate per symbol $K(\bar{n}; \tau,  n_\mathrm{n})$ can be given in a closed mathematical form\cite{LaudenbachAQT2018,PirandolaAOP2020}. When the channel parameters $\tau$ and $n_\mathrm{n}$ are fixed, the maximum key rate is obtained in the limit $\bar{n} \rightarrow \infty$, although in contrast to the information capacity the asymptotic value remains finite for the former\cite{BanaszekECOC2023}. 

\section{Pulse shaping}
In practice, the transmitter subsystem may have a limited capability to generate pulses in the required shape. This will be modelled by taking the signal generated by Alice in the form
\begin{equation}
    s(t) = \sum_{j=-\infty}^{\infty} \alpha_j u(t-j T), 
\end{equation}
where the normalized function $u(t)$ describes the actual generated pulse shape that does not necessarily satisfy the orthogonality condition specified in Eq.~(\ref{Eq:Orthogonality}). Assuming that the receiver implements detection in orthogonal modes $v(t-j'T)$, the complex amplitude detected by Bob for the $j'$th symbol is
\begin{equation}
\beta_{j'}= \int_{-\infty}^{\infty}  \mathrm{d}t \, v^\ast(t-j'T) s(t) = \sum_{j=-\infty}^{\infty} c_{j'-j} \alpha_j,
\label{Eq:betaalfa}
\end{equation}
where
\begin{equation}
    c_{j'} = \int_{-\infty}^{\infty} v^\ast(t-j'T) u(t). 
\end{equation}
Eq.~(\ref{Eq:betaalfa}) indicates a two-fold effect of mode mismatch on the performance of the QKD system. First, the optical energy effectively detected for the $j'$th symbol is subject to further reduction by a factor $|c_0|^2$. Second, ISI contributions from other symbols with $j \neq j'$ result in an additional noise with an average photon number $\tau\bar{n}\sum_{j\neq 0} |c_j|^2$. Consequently, the effective secure key rate SKR per symbol reads:
\begin{equation}
    \mbox{SKR} = K\bigl(\bar{n}; 
    \tau|c_0|^2,  n_\mathrm{n} + {\textstyle \tau\bar{n}\sum_{j\neq 0} |c_j|^2 }\bigr). 
    \label{Eq:SKR}
\end{equation}

\section{Results}

\begin{figure}[t!]
    \centering
    \includegraphics[width=7.5cm]{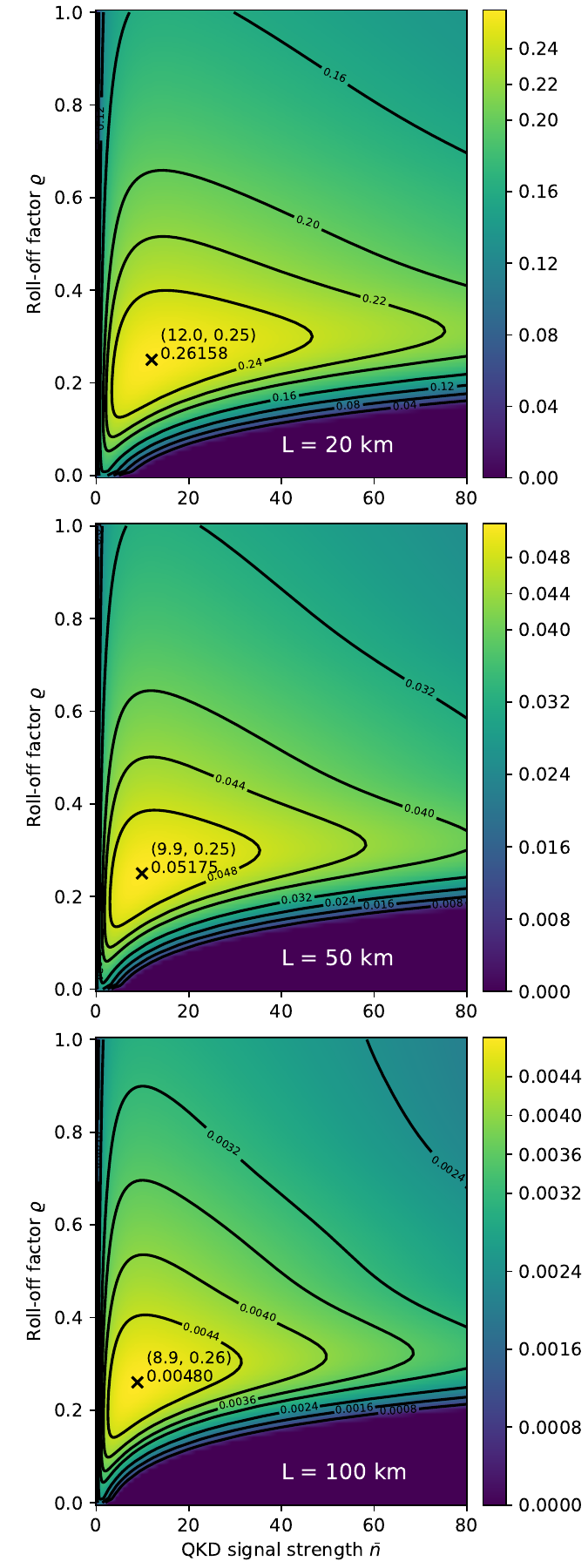}
    \caption{Key spectral efficiency KSE as a function of the input signal strength characterized by the average photon number per symbol $\bar{n}$ and the roll-off factor $\varrho$ of the RRC profile.}
    \label{Fig:KSE}
\end{figure}

The general model presented above has been applied to a QKD signal composed of RRC pulses approximated by step functions at 3 samples-per-symbol rate and truncated to 21 taps per symbol as shown in Fig.~\ref{Fig:RRCprofiles}. The relevant figure of merit is the key spectral efficiency $\mbox{KSE}=\mbox{SKR}/(1+\varrho)$ which takes into account the bandwidth overhead resulting from a non-zero value of the roll-off factor $\varrho$. Fig.~\ref{Fig:KSE} shows KSE as a function of the 
the average signal photon number $\bar{n}$ and the roll-off factor $\varrho$ for distances $20~\textrm{km}$, $50~\textrm{km}$, and $100~\textrm{km}$ assuming standard $0.2~\textrm{dB/km}$ attenuation.

Two observations are in place. First, while for perfectly matched Tx-Rx modes raising the signal strength monotonically increases the key rate, in the present case there is a well pronounced maximum for a finite photon number $\bar{n}$. This can be attributed to the fact that for mismatched modes increasing the signal strength enhances the ISI contribution to the excess noise, which outweighs the benefit of a higher signal strength. Second, one can identify the optimal value of the roll-off factor $\varrho$ that maximizes KSE. This optimum rather weakly depends on the signal strength or the channel transmission and stems from the interplay between detrimental effects of pulse truncation in the temporal domain that are more substantial for lower roll-offs, and the wider bandwidth occupied by a signal with a higher $\varrho$. 

\begin{figure}
    \centering
    \includegraphics[width=7.5cm]{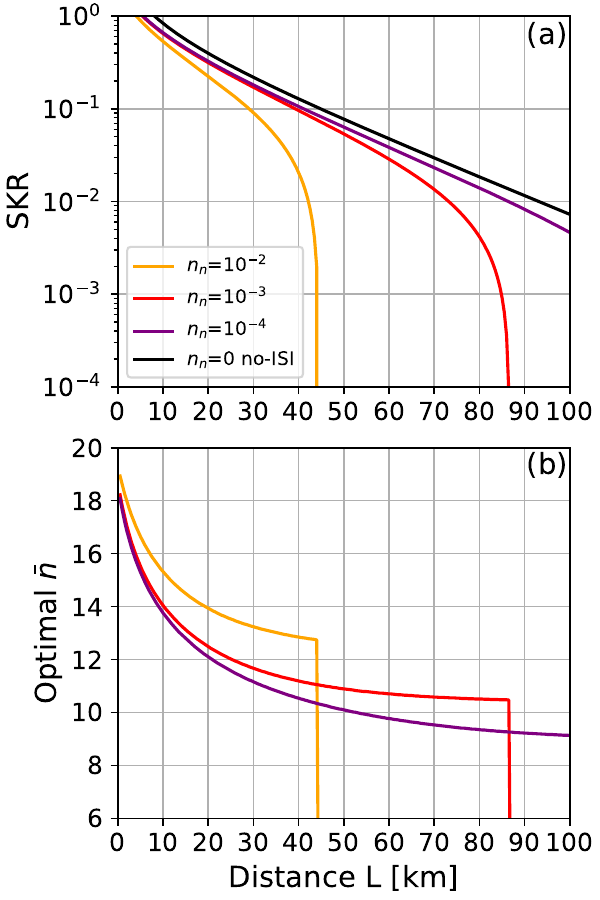}
    \caption{The optimal secret key rate (a) and the input signal strength (b) as a function of distance for non-zero excess noise $n_\mathrm{n} = 10^{-2}, 10^{-3}, 10^{-4}$ and the roll-off factor $\varrho=0.25$.}
    \label{Fig:SKRnn}
\end{figure}

Fig.~\ref{Fig:SKRnn}(a) depicts the attainable SKR for the roll-off factor $\varrho = 0.25$ as a function of distance taking into account mode-mismatch effects when the channel contributes non-zero excess noise $n_\mathrm{n}$ at the output. Qualitatively, dependence of the SKR on the distance is analogous to that attainable in the mode-matched scenario\cite{BanaszekECOC2023}, with a relatively minor penalty resulting from mode mismatch for short distances before the cliff region where effects of the excess noise set in. As seen in Fig.~\ref{Fig:SKRnn}(b), the optimal input signal strength $\bar{n}$ exhibits non-trivial dependence on channel parameters, but before the cliff is reached the attainable key rate depends rather weakly on this parameter, as indicated by  Fig.~\ref{Fig:SKRnn}(a) and also Fig.~\ref{Fig:KSE}. Importantly, for long distances that can be reached when the excess noise is low, the optimal signal strength becomes modest, falling well below 10 photons per symbol.

When the total number of taps is restricted, e.g.\ by the power consumption budget, and their temporal spacing is determined by the electronic bandwidth, one can consider transmission at different symbol rates defined by the number of samples per symbol (sps). Tab.~\ref{Tab:KSEsps} specifies the maximum attainable KSE and the corresponding optimal roll-off factors and the signal strengths for different sps choices with 21 samples in total. A clear trade-off between the quality of the step-function approximation of the RRC profile and truncation effects is clearly seen, indicating 3 sps as the optimum for the assumed pulse shaping scenario, independently of the transmission distance.

\begin{table}[h!]
    \centering
    \caption{The attainable key spectral efficiency KSE for different numbers of samples per symbol with the total of 21 samples.} \label{Tab:KSEsps}
    \begin{tabular}{|c|c|c|c|}
        \hline  Samples & Roll-off & Signal & KSE            \\
              per symbol & factor & strength &    \\
       \hline       \multicolumn{4}{|c|}{Distance $L=20~\textrm{km}$} \\
        \hline  2 & 0.15 & \phantom{0}6.6     &     0.24165    \\
        \hline \cellcolor{yellow}3 & \cellcolor{yellow}0.25 & \cellcolor{yellow}12.0       &    \cellcolor{yellow}0.26158    \\
        \hline  4 & 0.39 & 19.6        &   0.25318    \\
        \hline  6 & 0.64 & 51.7      &     0.23014    \\
        \hline  8 & 0.73 & 54.1       &    0.21954     \\
       \hline       \multicolumn{4}{|c|}{Distance $L=50~\textrm{km}$} \\
        \hline  2 & 0.16 & \phantom{0}5.6     &     0.04823    \\
        \hline  \cellcolor{yellow}3 & \cellcolor{yellow}0.25 & \cellcolor{yellow}\phantom{0}9.9       &    \cellcolor{yellow}0.05175     \\
        \hline  4 & 0.39 & 16.3        &   0.04986    \\
        \hline  6 & 0.64 & 43.7      &     0.04509    \\
       \hline  8 & 0.71 & 34.0       &    0.04298     \\
       \hline       \multicolumn{4}{|c|}{Distance $L=100~\textrm{km}$} \\
        \hline  2 & 0.16 & \phantom{0}4.6     &     0.00442    \\
        \hline  \cellcolor{yellow}3 & \cellcolor{yellow}0.26 &
        \cellcolor{yellow}\phantom{0}8.9       &    \cellcolor{yellow}0.00480     \\
        \hline  4 & 0.40 & 15.2       &   0.00465    \\
        \hline  6 & 0.64 & 38.2      &     0.00423    \\
        \hline  8 & 0.73 & 40.0       &    0.00403     \\
        \hline
    \end{tabular}
\end{table}

\section{Conclusions}
The mathematical model presented in this paper quantifies the effects of the mismatch between the CV QKD signal generated by a transmitter with a limited pulse shaping capability and the actual temporal modes detected by a coherent receiver. Such mismatch lowers the detected signal power and produces, via ISI, 
an additional contribution to the effective channel excess noise. For a transmitter generating discretized and truncated RRC pulses one can identify the optimal value of the roll-off factor that maximizes the key spectral efficiency. In contrast to the idealized mode-matched scenario, the maximum key rate is achieved for a finite input signal strength because of the growing detrimental ISI contribution to the excess noise. It can be expected that in the case of CV QKD systems based on discrete constellations \cite{Djordjevic2019, RoumestanECOC2021, Pereira2022, Wang2022, Wang2023}, this effect will limit the useful constellation size.

\clearpage

\section{Acknowledgements}
We acknowledge insightful discussions with K. {\L}ukanowski.
This work has been supported by the European
Union’s Horizon Europe research and innovation programme under the project ‘Quantum Security
Networks Partnership’ (QSNP, Grant Agreement No. 101114043).

\printbibliography

@ARTICLE{BanaszekJLT2020,
  author={Banaszek, Konrad and Kunz, Ludwig and Jachura, Michał and Jarzyna, Marcin},
  journal={J. Lightwave Technol.}, 
  title={Quantum Limits in Optical Communications}, 
  year={2020},
  volume={38},
  pages={2741-2754},
  doi={10.1109/JLT.2020.2973890},}

@article{PirandolaAOP2020,
author = {S. Pirandola and U. L. Andersen and L. Banchi and M. Berta and D. Bunandar and R. Colbeck and D. Englund and T. Gehring and C. Lupo and C. Ottaviani and J. L. Pereira and M. Razavi and J. Shamsul Shaari and M. Tomamichel and V. C. Usenko and G. Vallone and P. Villoresi and P. Wallden},
journal = {Adv. Opt. Photon.},
pages = {1012--1236},
publisher = {Optica Publishing Group},
title = {Advances in quantum cryptography},
volume = {12},
year = {2020},
doi = {10.1364/AOP.361502},
}

@Article{LaudenbachAQT2018,
  author    = {Fabian Laudenbach and Christoph Pacher and Chi-Hang Fred Fung and Andreas Poppe and Momtchil Peev and Bernhard Schrenk and Michael Hentschel and Philip Walther and Hannes Hübel},
  journal   = {Advanced Quantum Technologies},
  title     = {Continuous-Variable Quantum Key Distribution with {G}aussian Modulation-The Theory of Practical Implementations},
  year      = {2018},
  pages     = {1800011},
  volume    = {1},
  doi       = {10.1002/qute.201800011},
  publisher = {Wiley},
}

@article{PanWangOL2022,
author = {Yan Pan and Heng Wang and Yun Shao and Yaodi Pi and Yang Li and Bin Liu and Wei Huang and Bingjie Xu},
journal = {Opt. Lett.},
number = {13},
pages = {3307--3310},
publisher = {Optica Publishing Group},
title = {Experimental demonstration of high-rate discrete-modulated continuous-variable quantum key distribution system},
volume = {47},
year = {2022},
doi = {10.1364/OL.456978},
}

@INPROCEEDINGS{RoumestanECOC2021,
  author={Roumestan, François and Ghazisaeidi, Amirhossein and Renaudier, Jérémie and Vidarte, Luis Trigo and Diamanti, Eleni and Grangier, Philippe},
  booktitle={2021 European Conference on Optical Communication (ECOC)}, 
  title={High-Rate Continuous Variable Quantum Key Distribution Based on Probabilistically Shaped 64 and 256-{QAM}}, 
  year={2021},
  volume={},
  number={},
  pages={1-4},
  doi={10.1109/ECOC52684.2021.9606013},
  ISSN={},}

@INPROCEEDINGS{BanaszekECOC2023,
  author={Banaszek, K. and Łukanowski, K. and Jachura, M. and Jarzyna, M.},
  booktitle={49th European Conference on Optical Communications (ECOC 2023)}, 
  title={Quantum noise in optical communication systems: limitations and opportunities}, 
  year={2023},
    number={},
  pages={349-352},
  doi={10.1049/icp.2023.2077},
  ISSN={},}

@ARTICLE{Armstrong2009,
  author={Armstrong, Jean},
  journal={Journal of Lightwave Technology}, 
  title={{OFDM} for Optical Communications}, 
  year={2009},
  volume={27},
  number={3},
  pages={189-204},
  doi={10.1109/JLT.2008.2010061}}

@INPROCEEDINGS{RoumestanOFC2021,
  author={Roumestan, François and Ghazisaeidi, Amirhossein and Renaudier, Jérémie and Brindel, Patrick and Diamanti, Eleni and Grangier, Philippe},
  booktitle={2021 Optical Fiber Communications Conference and Exhibition (OFC)}, 
  title={Demonstration of Probabilistic Constellation Shaping for Continuous Variable Quantum Key Distribution}, 
  year={2021},
  volume={},
  number={},
  pages={1-3},
  doi={},
  ISSN={},}

@Article{Hajomer2022,
author={Hajomer, A. A. E. and Jain, N. and Mani, H. and Chin, H. and Andersen, U. L. and Gehring, T.},
journal={npj Quantum Information},
title={Modulation leakage-free continuous-variable quantum key distribution},
year={2022},
volume={8},
pages={136},
doi={10.1038/s41534-022-00640-1},
}

@article{Hao2016,
  title = {Quantum hacking: Saturation attack on practical continuous-variable quantum key distribution},
  author = {Qin, Hao and Kumar, Rupesh and All\'eaume, Romain},
  journal = {Phys. Rev. A},
  volume = {94},
  pages = {012325},
  numpages = {16},
  year = {2016},
  publisher = {American Physical Society},
  doi = {10.1103/PhysRevA.94.012325},
}

@article{Jouguet2013,
  title = {Preventing calibration attacks on the local oscillator in continuous-variable quantum key distribution},
  author = {Jouguet, Paul and Kunz-Jacques, S\'ebastien and Diamanti, Eleni},
  journal = {Phys. Rev. A},
  volume = {87},
  pages = {062313},
  numpages = {6},
  year = {2013},
  publisher = {American Physical Society},
  doi = {10.1103/PhysRevA.87.062313},
}

@ARTICLE{Djordjevic2019,
  author={Djordjevic, Ivan B.},
  journal={IEEE Access}, 
  title={On the Discretized Gaussian Modulation ({DGM})-Based Continuous Variable-{QKD}}, 
  year={2019},
  volume={7},
  number={},
  pages={65342-65346},
  doi={10.1109/ACCESS.2019.2917587}}

@article{Pereira2022,
author = {Daniel Pereira and Margarida Almeida and Margarida Fac\~{a}o and Armando N. Pinto and Nuno A. Silva},
journal = {Opt. Lett.},
pages = {3948--3951},
publisher = {Optica Publishing Group},
title = {Probabilistic shaped 128-{APSK} {CV}-{QKD} transmission system over optical fibres},
volume = {47},
year = {2022},
doi = {10.1364/OL.456333},
}

@article{Wang2023,
doi = {10.1088/1367-2630/acb964},
year = {2023},
publisher = {IOP Publishing},
volume = {25},
pages = {023019},
author = {Pu Wang and Yu Zhang and Zhenguo Lu and Xuyang Wang and Yongmin Li},
title = {Discrete-modulation continuous-variable quantum key distribution with a high key rate},
journal = {New Journal of Physics},
}

@article{Wang2022,
  author    = {Heng Wang and
               Yang Li and
               Yaodi Pi and
               Yan Pan and
               Yun Shao and
               Li Ma and
               Yichen Zhang and
               Jie Yang and
               Tao Zhang and
               Wei Huang and
               Bingjie Xu},
  title     = {Sub-{G}bps key rate four-state continuous-variable quantum key distribution within metropolitan area},
  journal   = {Communications Physics},
  year      = {2022},
  volume    = {5},
  pages     = {162},
  issn      = {2399-3650},
  doi       = {10.1038/s42005-022-00941-z},
}

\vspace{-4mm}

\end{document}